\documentclass[]{aa}
\usepackage{graphicx,natbib}

\def\ms{\mbox{$M_\odot$}}
\def\ds{\mbox{$\rm d_\odot$}}
\def\dgc{\mbox{$R_{\rm GC}$}}
\def\kms{\mbox{$\rm km\,s^{-1}$}}
\def\msy{\mbox{$\rm mas\,yr^{-1}$}}

\begin{document}

\title{On the possible generation of the young massive open clusters 
Stephenson\,2 and BDSB\,122 by $\omega$\,Centauri}

\author{G.M. Salerno\inst{1} \and E. Bica\inst{1} \and C. Bonatto\inst{1} 
\and I. Rodrigues\inst{2}}

\offprints{C. Bonatto}

\institute{Universidade Federal do Rio Grande do Sul, Departamento de Astronomia\\
CP\,15051, RS, Porto Alegre 91501-970, Brazil\\
\email{salerno@if.ufrgs.br, bica@if.ufrgs.br, charles@if.ufrgs.br, irapuan@univap.br}
\mail{charles@if.ufrgs.br}\\
\and
IP\&D - Universidade do Vale do Para\'\i ba - UNIVAP, Av. Shishima Hifumi, 2911 - Urbanova,
S\~ao Jos\'e dos Campos 12244-000, SP, Brazil }

\date{Received --; accepted --}

\abstract
{A massive objects such as a globular cluster passing through the disk of a galaxy 
can trigger star formation.}
{We test the hypothesis that the most massive globular cluster in the Galaxy, 
$\omega$\,Centauri, which crossed the disk approximately $24\pm2$\,Myr ago, may
have triggered the formation of the open clusters Stephenson\,2 and BDSB\,122. }
{The orbits of $\omega$\,Centauri, Stephenson\,2 and BDSB\,122 are computed for 
the three-component model of Johnston, Hernquist \& Bolte, which considers the 
disk, spheroidal and halo gravitational potentials. }
{With the re-constructed orbit of $\omega$\,Centauri, we show that the latest impact 
site is consistent, within important uncertainties, with the birth-site of the young massive open clusters 
BDSB\,122 and Stephenson\,2. Within uncertainties, this scenario is consistent with 
the time-scale of their backwards motion in the disk, shock wave propagation and delay 
for star formation. }
{Together with open cluster formation associated to density waves in spiral arms, the 
present results are consistent with the idea that massive globular clusters as additional progenitors of open 
clusters, the massive ones in particular.}


\titlerunning{Massive open clusters generated by $\omega$\,Centauri?}

\maketitle

\section{Introduction}
\label{intro}

Disk-stability criteria and impact assumptions suggest that the passage of a globular cluster
(GC) can trigger a bubble or wave of self-propagating star formation within the disk of the Galaxy 
(\citealt{Wallin96}). The initial mechanical perturbation produces a local enhancement of the 
interstellar medium (ISM) density, from which localised star formation may occur. Subsequently,
clustered star formation may happen along the border of a radially expanding density wave or 
ionisation front (e.g. \citealt{Soria05} -- hereafter SCP05; \citealt{Elme77}; \citealt{Whit94}). 
The expanding bubble is capable
of compressing the neutral ISM above the stability criterion against gravitational collapse. 
Alternative star-formation triggering mechanisms are the infall of a high-velocity HI cloud 
(\citealt{Elme00}; \citealt{Larsen02}), or hypernova explosions (\citealt{Pacz98}).
 
Prominent, isolated star-forming bubbles have been observed in external galaxies. A
bubble with $\approx600$\,pc in diameter was detected in NGC\,6946 (\citealt{Larsen02}), 
containing a young super star cluster and at least 12 surrounding young clusters, the latter being 
comparable to the most luminous Galactic OCs. The triggering mechanism in NGC\,6946 
appears to be the impact of a high-velocity HI cloud and/or hypernova explosions (\citealt{Elme00}). 
The Galaxy may harbour similar structures. A possible example is the Cygnus superbubble,
which contains OB associations (\citealt{Vlemmings04}, and references therein). 

For a GC, the triggering effects are essentially gravitational. A natural
assumption is that GCs, crossing the disk every 1\,Myr on average, may be responsible 
for some star formation. A possible case relates the origin of the 
OC NGC\,6231 to the GC NGC\,6397 disk-crossing (\citealt{Rees03}). Another possibility is
the low-mass GC FSR\,584 as star-formation trigger in the W\,3 complex (\citealt{FSR584}).

\begin{figure}
\begin{minipage}[b]{0.5\linewidth}
\includegraphics[width=\textwidth]{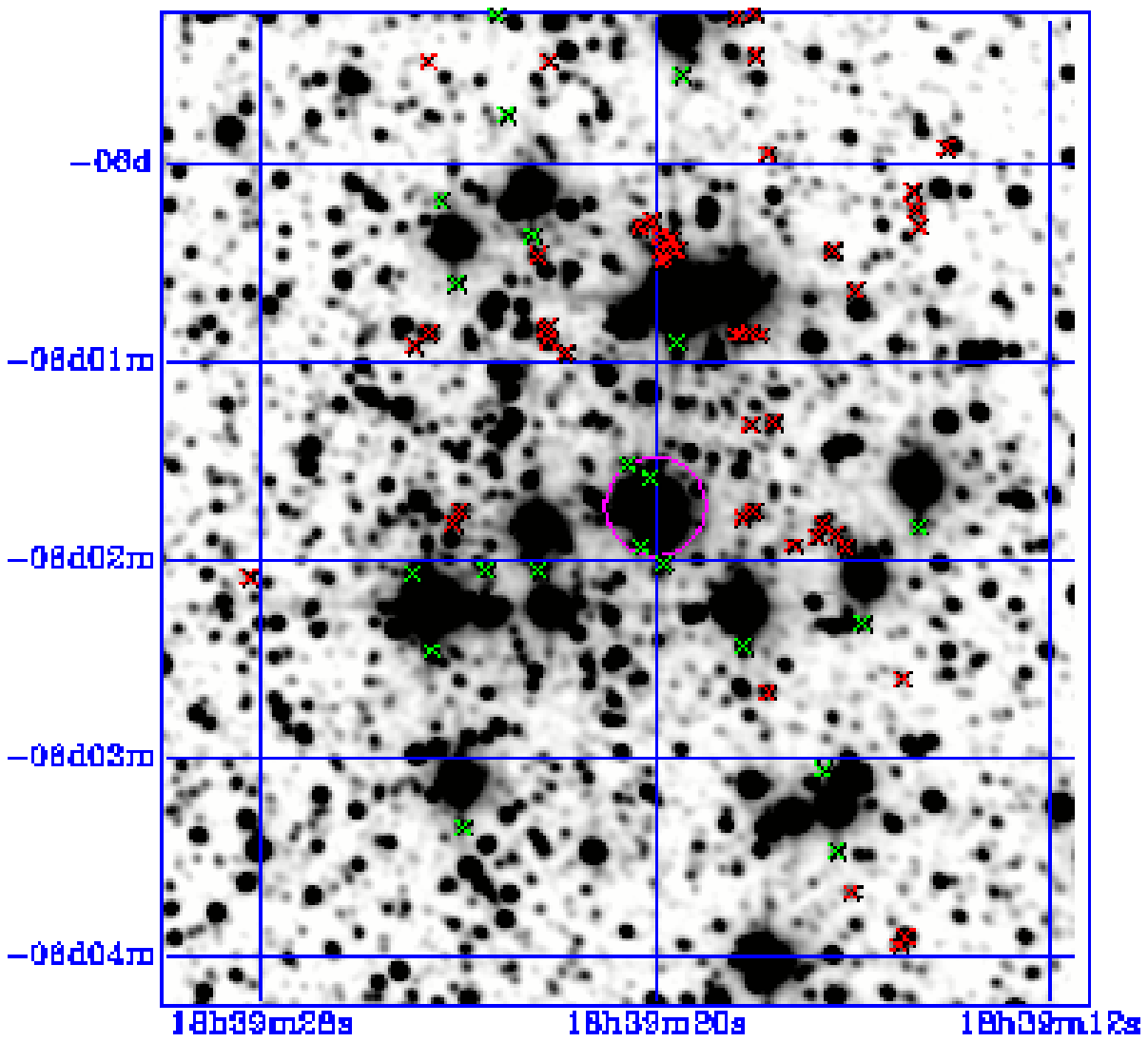}
\end{minipage}\hfill
\begin{minipage}[b]{0.5\linewidth}
\includegraphics[width=\textwidth]{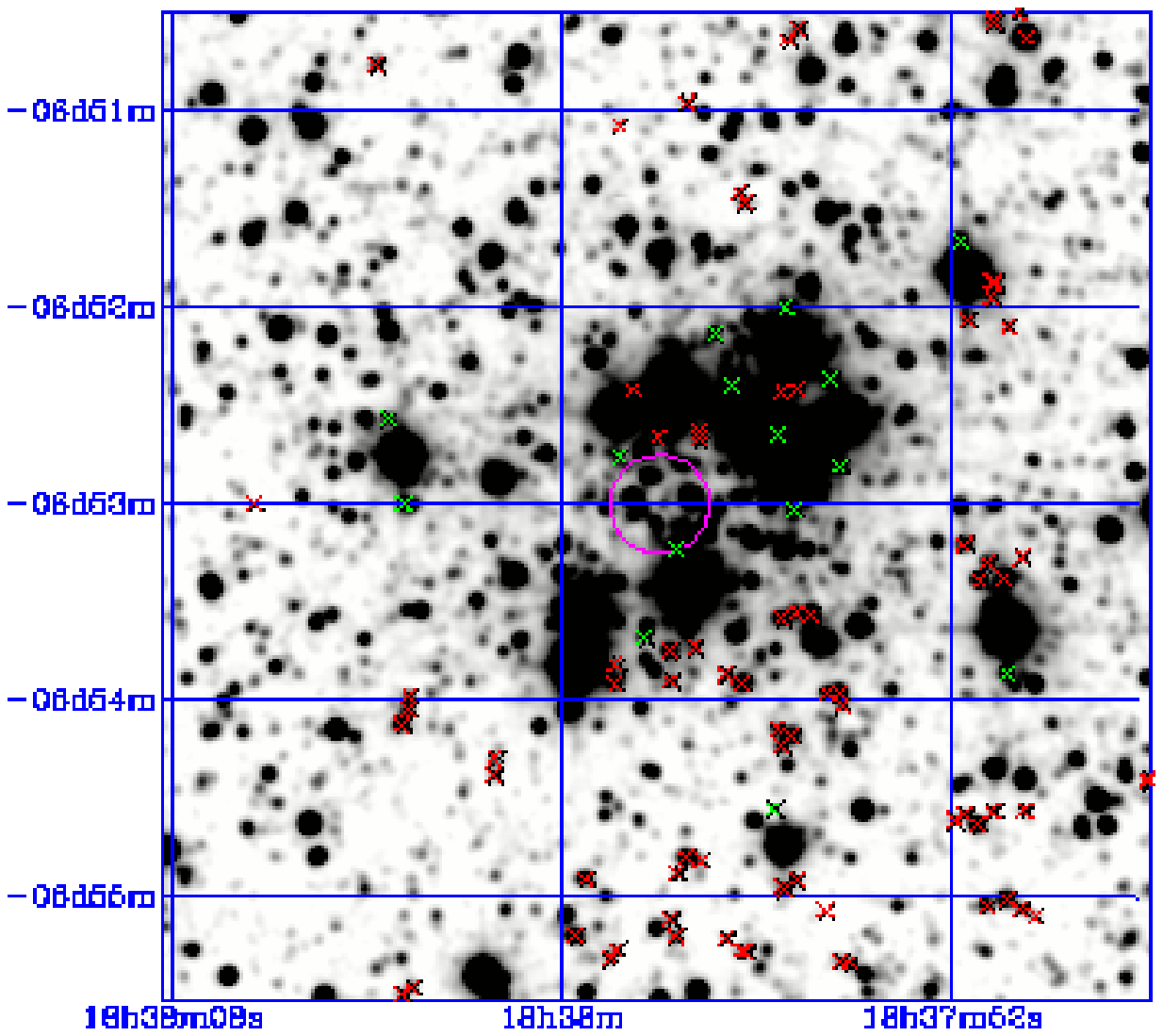}
\end{minipage}\hfill
\caption[]{$5\arcmin\times5\arcmin$ 2MASS $K_S$ images of Stephenson\,2 (left) 
and BDSB\,122 (right). Figure orientation: North to the top and East to the left.}
\label{fig1}
\end{figure}

The OCs Stephenson\,2 and BDSB\,122 were discovered in 1990 (\citealt{Steph90}) and 2003
(\citealt{BDSB03}), respectively. 2MASS ({\em
www.ipac.caltech.edu/2mass/releases/allsky}) images of both clusters are shown in Fig.~\ref{fig1}. 
The suspected richness of Stephenson\,2 in red supergiants was
confirmed by \citet{Nakaya01} and \citet{OBBM02}, providing an age of $\approx20$\,Myr, and a distance
from the Sun $\ds=6$\,kpc (\citealt{OBBM02}). Both clusters are among the most massive OCs
known in the Galaxy. Indeed, BDSB\,122 has 14 red supergiants, is located at $\ds=5.8$\,kpc from the Sun, 
an estimated mass of $2-4\times10^4\,\ms$, and an age of $7-12$\,Myr (\citealt{Figer06}). Stephenson\,2 
has 26 red supergiants, is located at $\ds=5.8^{+1.9}_{-0.8}$\,kpc from the Sun, has an estimated mass 
of $4\times10^4\,\ms$, and an age of $12-17$\,Myr (\citealt{Davies07}). Their distances from the Sun are the 
same, within uncertainties, and their projected separation on the sky is $\approx100$\,pc. The designation 
Stephenson\,2 was originally given by \citet{OBBM02}, and also adopted by Dias et al. (2002, and updates).
Stephenson\,2 and BDSB\,122 are clearly in the red supergiant 
(RSG) phase (\citealt{BSA90}). \citet{Davies07} refer to these clusters as RSGC\,1 and RSGC\,2, respectively. 

\begin{figure}
\resizebox{\hsize}{!}{\includegraphics{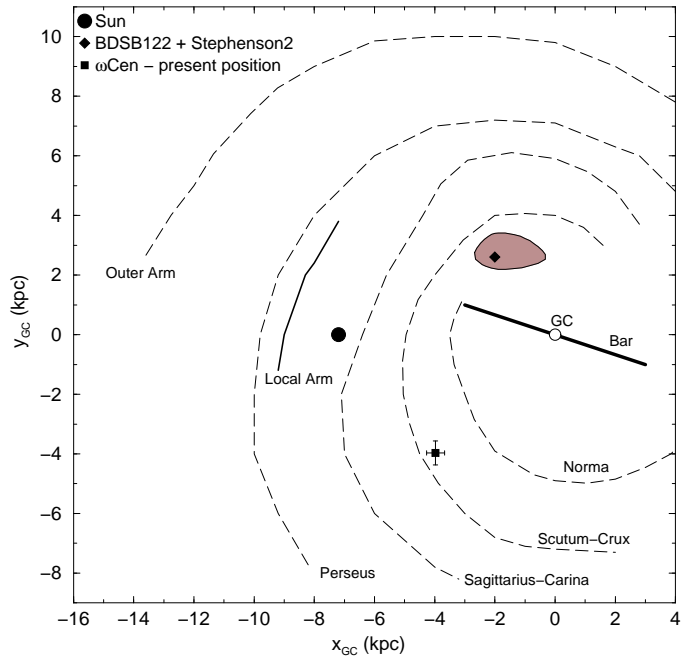}}
\caption[]{The present-day positions (and uncertainties) of Stephenson\,2, 
BDSB\,122, and $\omega$\,Centauri overplotted on a schematic projection of the Galaxy as seen 
from the North pole, with 7.2\,kpc as the Sun's distance to the Galactic centre. Main structures 
are identified.}
\label{fig1b}
\end{figure}

The positions (and uncertainties) of both clusters, together with $\omega$ Centauri (NGC\,5139), are shown
in Fig.~\ref{fig1b} superimposed on a schematic view of the Milky Way (based on \citealt{Momany06} and
\citealt{DrimSpe01}). The locus of BDSB\,122 and Stephenson\,2 is slightly more internal 
than the Scutum-Crux arm. Several other young clusters from the catalogues of \citet{BDSB03} and
\citet{DBSB03} have already been studied in detail (e.g. \citealt{SBAC08}; \citealt{OBBMB08}; \citealt{HB08}).

We trace backwards in time the orbits of $\omega$ Centauri and that of 
the OC Stephenson\,2 (and consequently, also that of BDSB\,122) in the disk, testing an impact
hypothesis for the origin of these two massive OCs. Orbit integrations in the Galactic potential 
using as constraints the GC space velocity have been widely applied to 54 GCs
(e.g. \citealt{Dinescu03}; \citealt{AMP08}).

This paper is structured as follows. In Sect.~\ref{W100} we study the past orbit of
$\omega$\,Centauri. In Sect.~\ref{Orbit}, the past orbits of Stephenson\,2 and 
$\omega$\,Centauri are compared looking for spatial and time coincidence. Conclusions 
are in Sect.~\ref{Conclu}.
 
\section{$\omega$\,Centauri as a projectile}
\label{W100}

$\omega$\,Centauri, the most massive Galactic GC ($4\times10^6\,\ms$- \citealt{Nakaya01}), has a
metallicity spread and a flat density distribution typical of a dwarf galaxy nucleus captured by
the Galaxy (\citealt{BekkiFree03}). Thus, irrespective of the existence of young massive clusters,
somehow associated with the impact site, the orbit of $\omega$\,Centauri under the Galactic potential
is worth studying, especially to look for consequences of the last disk passage.
Evidence of a similar disk impact and a star-forming event has been observed in the spiral galaxy
NGC\,4559 with Hubble Space Telescope (HST), XMM-Newton, and ground-based (SCP05) data.
The age of the star-forming complex is $\la30$\,Myr, with a ring-like distribution. It appears to
be an expanding wave of star formation, triggered by an initial density perturbation. The most likely
triggering mechanism was a collision with a satellite dwarf galaxy crossing through the gas-rich
outer disk of NGC\,4559, which appears to be the dwarf galaxy visible a few arcsec NW of the
complex. The picture is reminiscent of a scaled-down version of the Cartwheel galaxy
(\citealt{StruckM93}; \citealt{Struck96}).

As another example, proper motions (PMs) and radial velocity suggest that the GC
NGC\,6397 crossed the Galactic disk 5\,Myr ago, which possibly triggered the formation of the
OC NGC\,6231 (\citealt{Rees03}), thus lending support to the present scenario (\citealt{Wallin96}). 
NGC\,6397 and NGC\,6231 are closely projected on the sky ($\Delta\ell\approx5\degr$, $\Delta\,b\approx13\degr$).
However, in the case of $\omega$\,Centauri as possible generator of BDSB\,122 and Stephenson\,2,
the GC is now widely apart from the pair of massive OCs ($\Delta\ell\approx77\degr$,
$\Delta\,b\approx15\degr$). Thus, PM and radial velocity are fundamental constraints
for the analysis of $\omega$\,Centauri, and impact solutions require a detailed integration of
the orbit across the Galactic potential.

\begin{table}
\caption[]{Present-day cluster positions}
\label{tab1}
\renewcommand{\tabcolsep}{2.5mm}
\renewcommand{\arraystretch}{1.25}
\begin{tabular}{lrrrr}
\hline\hline
Cluster&$\ell$&$b$&$\alpha(J2000)$&$\delta(J2000)$ \\
&(\degr)&(\degr)&(h:m:s)&($\degr,\arcmin,\arcsec$)\\
(1)&(2)&(3)&(4)&(5)\\
\hline
$\omega$\,Centauri & 309.10 & $+$14.97 & 13:26:46 & $-47$:28:37\\
BDSB\,122          &  26.84 &  $+$0.65 & 18:37:58 & $-$6:53:00\\
Stephenson\,2      &  26.18 &  $-$0.06 & 18:39:20 & $-$6:01:44\\
\hline
\end{tabular}
\end{table}

\begin{figure*}
\begin{minipage}[b]{0.50\linewidth}
\includegraphics[width=\textwidth]{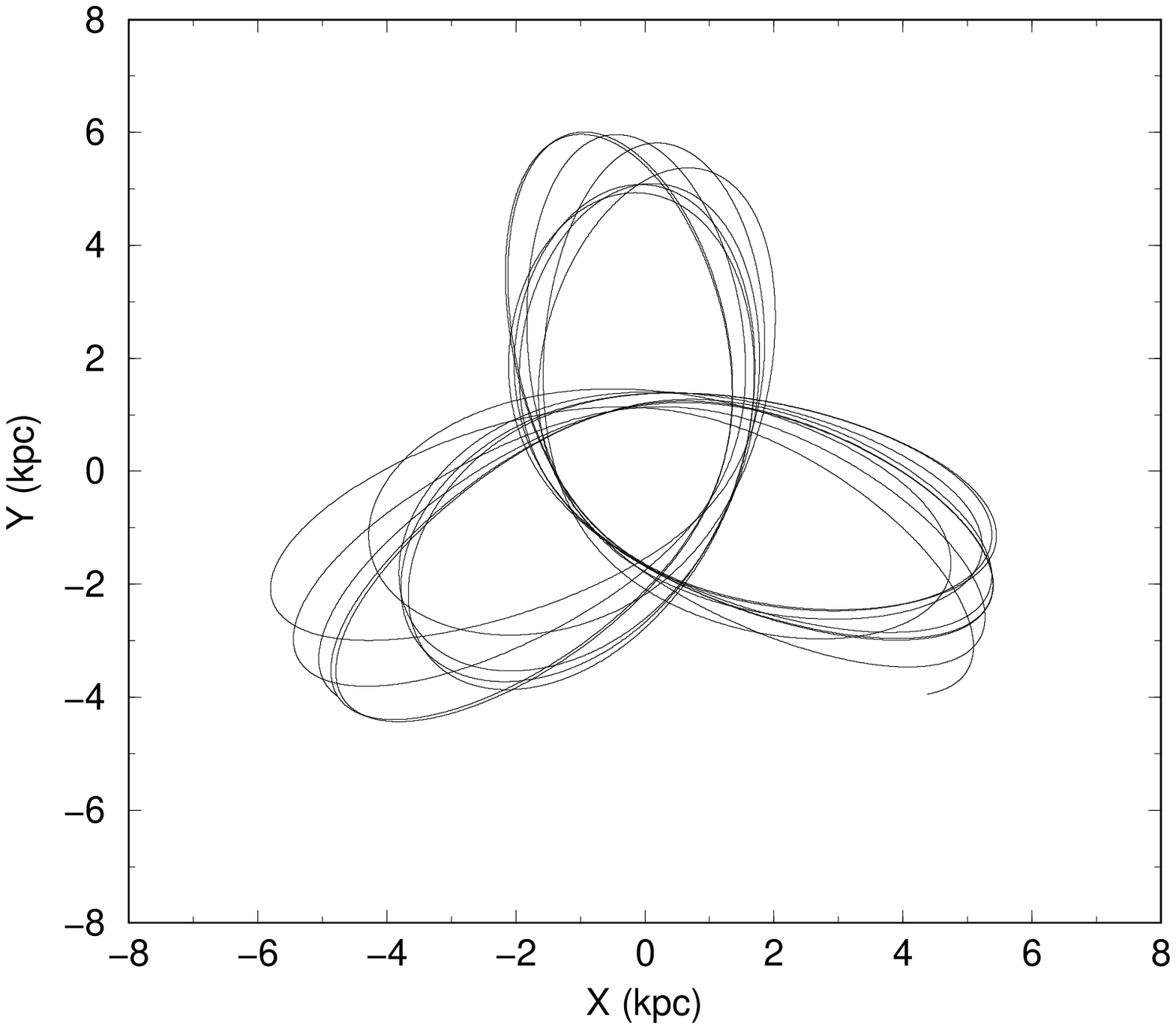}
\end{minipage}\hfill
\begin{minipage}[b]{0.50\linewidth}
\includegraphics[width=\textwidth]{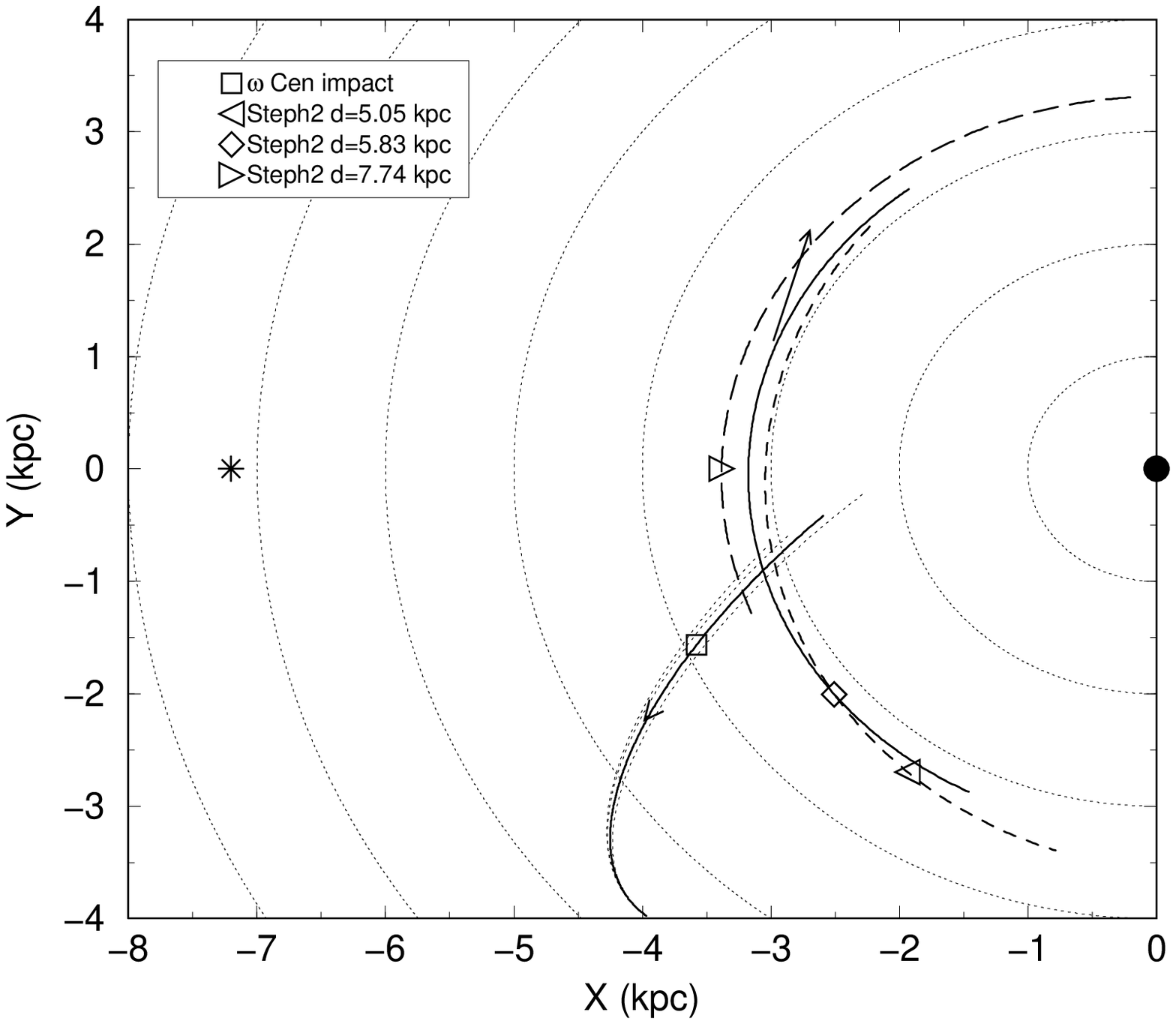}
\end{minipage}\hfill
\begin{minipage}[b]{0.50\linewidth}
\includegraphics[width=\textwidth]{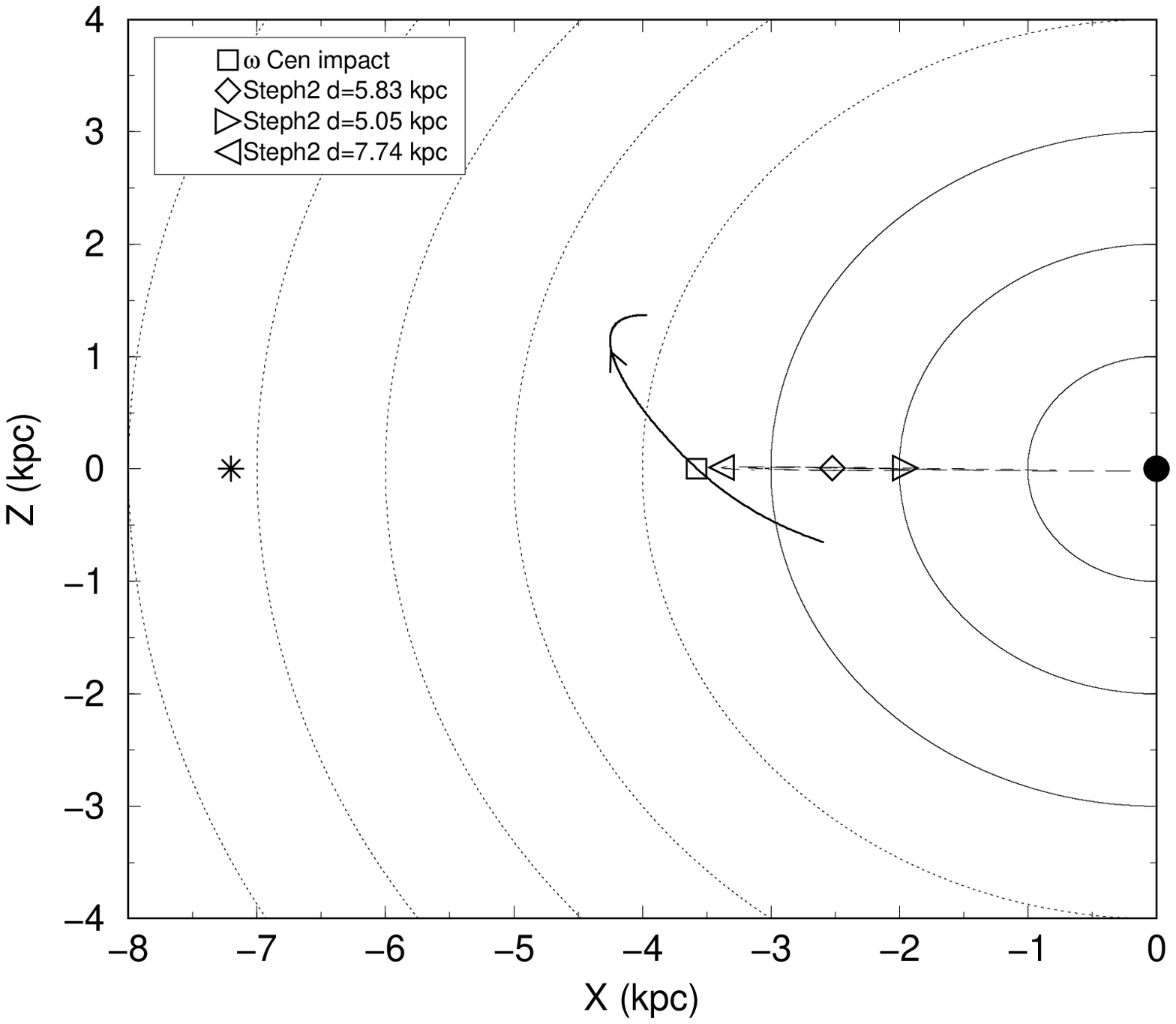}
\end{minipage}\hfill
\begin{minipage}[b]{0.50\linewidth}
\includegraphics[width=\textwidth]{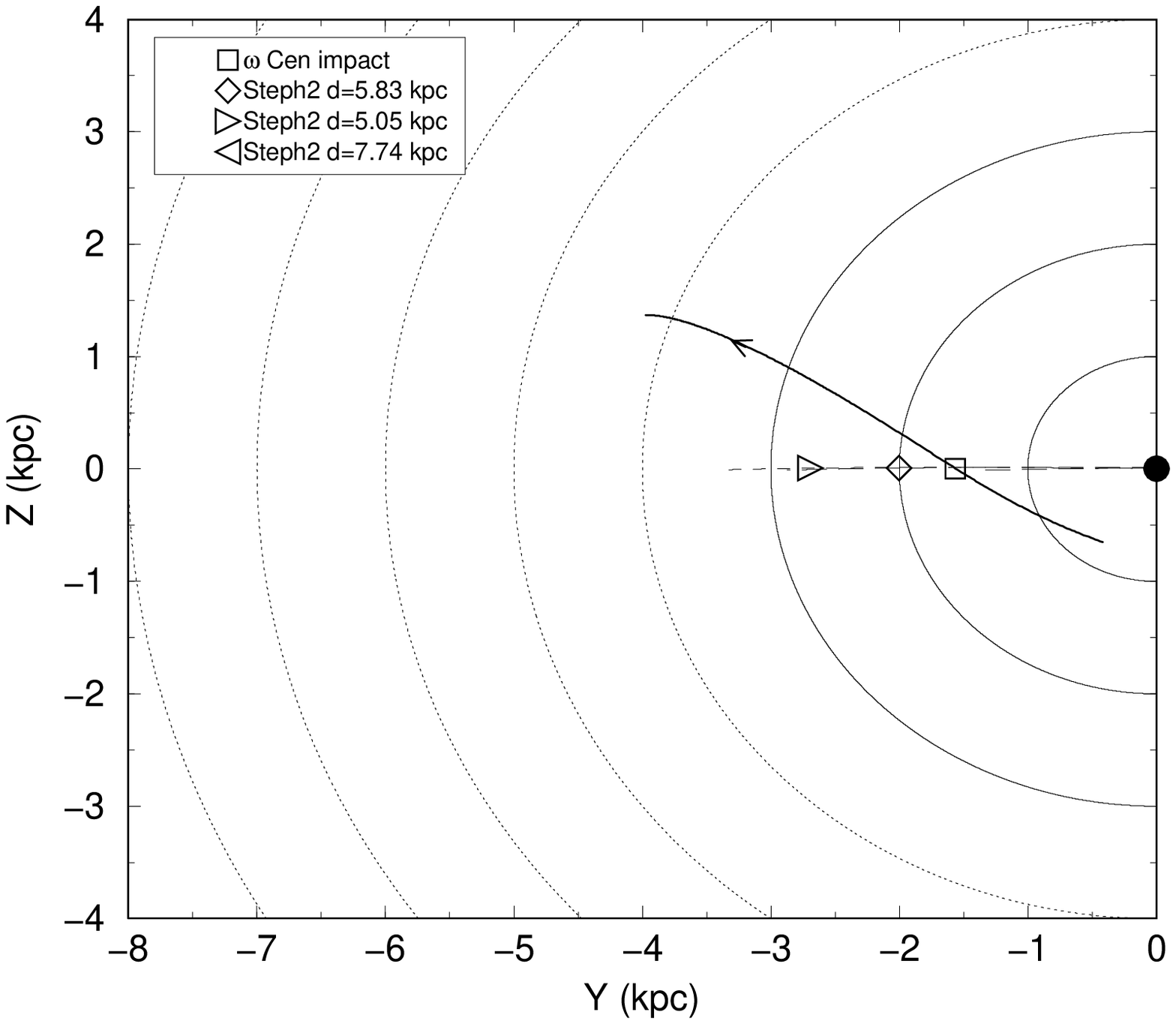}
\end{minipage}\hfill
\caption[]{Top-left panel: Galactocentric XY-plane projection of the $\omega$\,Centauri orbit over 
the past 2\,Gyr. Top-right: The past 30\,Myr orbit of $\omega$\,Centauri (solid line) for $\dgc=7.2$\,kpc. 
Additional neighbouring orbits (dotted) are, from bottom to top: JHB96 (-10\%), JHB96 (+10\%), FSC96.
The impact site on the disk is shown by the empty square. Arrows indicate orbit direction. Orbits of
Stephenson\,2 for the assumed distance from the Sun (and uncertainties) are shown. The corresponding
XZ and YZ projections are in the bottom panels. Empty symbols over the Stephenson\,2 orbits indicate 
its possible positions 24\,Myr ago. The Sun at its present position (asterisk) and the Galactic Centre 
(filled circle) are shown.}
\label{fig2}
\end{figure*}

\subsection{Orbit computation}
\label{Orbit}

The present three-component mass-distribution model of the Galaxy follows that employed in 
the study of a high-velocity black hole on a Galactic-halo orbit in the solar neighbourhood 
(\citealt{Mirabel01}, and references therein). In short, we use the three-component model
of \citet{Johnston96} -- hereafter JHB96 -- in which the disk, spheroidal, and halo gravitational potentials are
described as $\phi_{\rm disk}(R,z)=-GM_{\rm disk}/\sqrt{R^2+\left(a+\sqrt{z^2+b^2}\right)^2}$ 
(\citealt{MN75}), $\phi_{\rm spher}(R)=-\frac{GM_{\rm spher}}{R+c}$ (\citealt{Hern90}),
and $\phi_{\rm halo}(R)=v^2_{\rm halo}\ln{(R^2+d^2)}$, where $M_{\rm disk}=1\times10^{11}\,\ms$,
$M_{\rm spher}=3.4\times10^{10}\,\ms$, $v_{\rm halo}=128\,\kms$, $R$ and $z$ are the cylindrical 
coordinates, and the scale lengths $a=6.5$\,kpc, $b=0.26$\,kpc, $c=0.7$\,kpc, and $d=12.0$\,kpc.
Table~\ref{tab1} shows the Galactic and Equatorial coordinates of the three clusters.
Following \citet{Mizutani03}, the relevant parameters for computing the motion of 
$\omega$\,Centauri are the distance from the Sun $\ds=5.3\pm0.5$\,kpc, the PM 
components (\msy) $\mu_\alpha\cos(\delta)=-5.08\pm0.35$ and $\mu_\delta=-3.57\pm0.34$,
and finally the heliocentric radial velocity $V_r=232.5\pm0.7\,\kms$.

\begin{figure*}
\begin{minipage}[b]{0.50\linewidth}
\includegraphics[width=\textwidth]{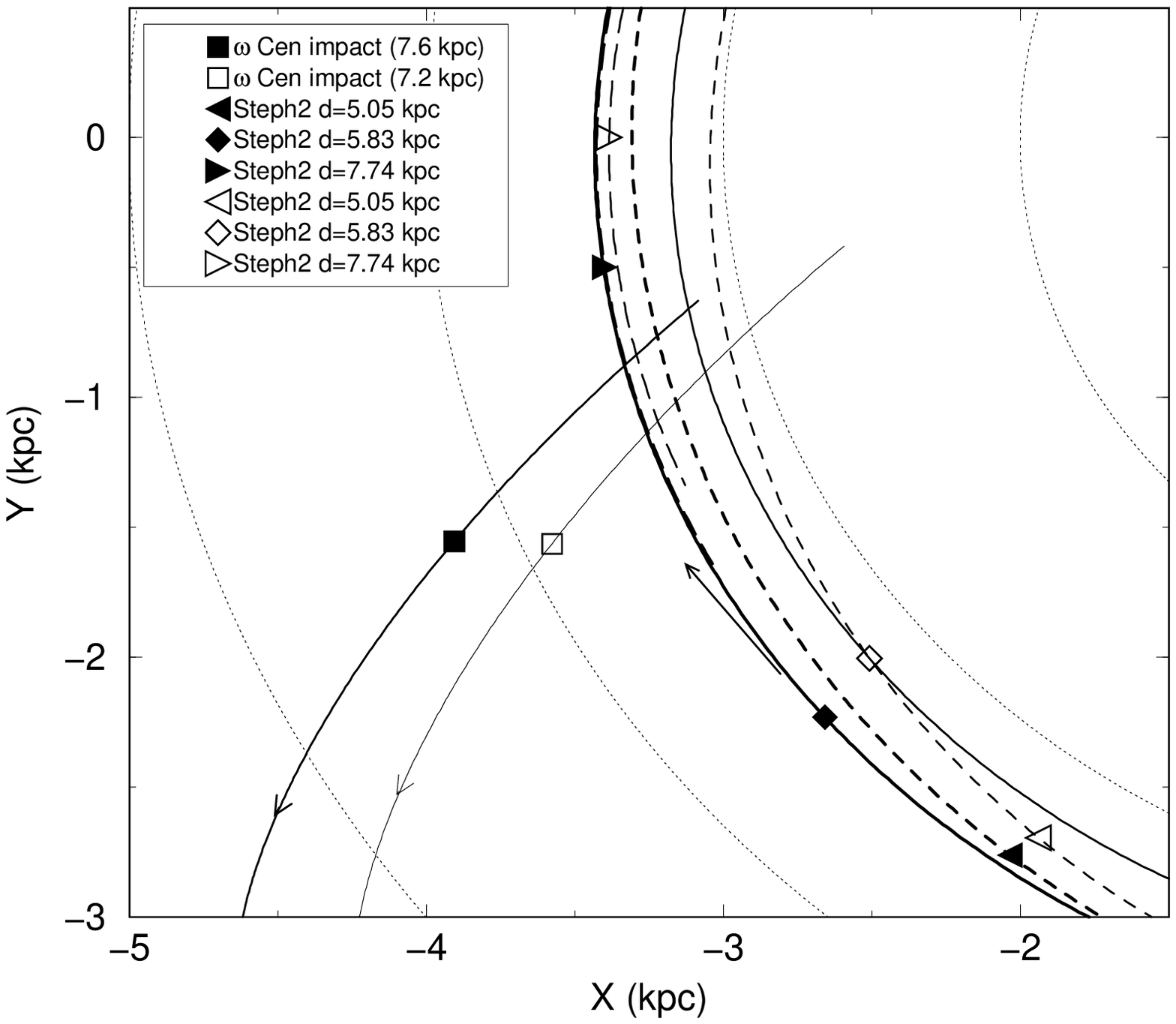}
\end{minipage}\hfill
\begin{minipage}[b]{0.50\linewidth}
\includegraphics[width=\textwidth]{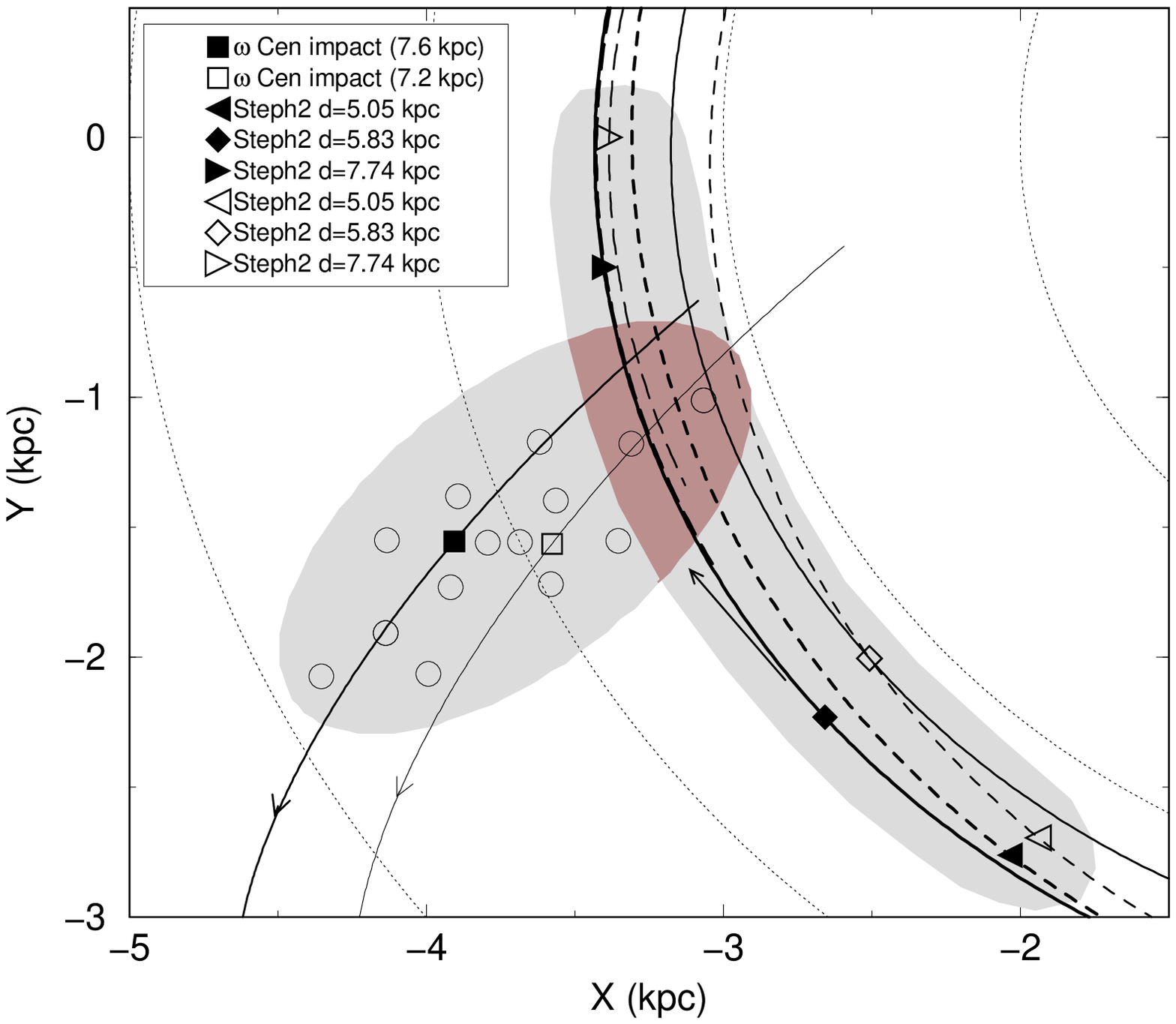}
\end{minipage}\hfill
\caption[]{Close-up of the impact site. Left panel: Orbits computed with $\dgc=7.2$\,kpc 
(empty symbols) and $\dgc=7.6$\,kpc (filled symbols). Right: Same as left panel but
including error distribution for the $\omega$\,Centauri impact site and Stephenson\,2 
proto-cluster position. A random selection of impact sites (open circles) is shown within 
the $\omega$\,Centauri error ellipsoid.}
\label{fig3}
\end{figure*}

The models were computed with $\dgc=7.2$\,kpc (\citealt{GCProp}) as the distance of the Sun 
to the Galactic centre. The Galactic velocities of $\omega$\,Centauri are $U=54.3\pm9.5\,\kms$, 
$V=-44.2\pm8.2\,\kms$, and $W=-1.3\pm13.0\kms$. Alternatively, we also computed orbits with
$\dgc=7.6$\,kpc, obtained by \citet{Eisenhauer05}. It should be noted that recently, by 
means of statistical parallax of central stars, \citet{Trippe08} found $\dgc=8.07\pm0.32$\,kpc,
while \citet{Ghez08}, with the orbit of one star close to the black hole, found $\dgc=8.0\pm0.6$\,kpc 
or $\dgc=8.4\pm0.4$\,kpc, under different assumptions. Cluster distances are heliocentric, which
do not depend on \dgc; on the other hand, the value of \dgc\ has some effect on the potentials,
which thus, affects orbit computation. Since the difference between the adopted value of \dgc\ and
the more recent ones is not exceeding, the value of \dgc\ should not influence much the present 
results.

Based on the rotation curves of \citet{BrBl93} and \citet{Russ03}, and an estimate with
the galaxy mass model described above (\citealt{Mirabel01}), we derived $V_c=214\pm4\,\kms$ as the orbital
velocity of Stephenson\,2. The nearly flat Galactic rotation curve at the Stephenson\,2 position 
allows to adopt this circular velocity also for the orbits corresponding to distance
uncertainties (Sect.~\ref{intro}). The orbit of $\omega$\,Centauri, computed back over 2\,Gyr,
is comparable to that derived by \citet{Mizutani03}, in particular the Rosette pattern seen projected 
on the XY plane (Fig.~\ref{fig2}). The simulation shows that $\omega$\,Centauri hit the disk as
recently as $24\pm2$\,Myr ago. Because of such a short time, fossil remains of this event may be
detectable in the disk nowadays.

Figure~\ref{fig2} (top-left panel) shows the Galactic XY-plane projection of the last 2\,Gyr
orbital motion of $\omega$\,Centauri. In the remaining panels we focus on the last 30\,Myr of
the motion of $\omega$\,Centauri and Stephenson\,2. For Stephenson\,2 we consider the
different orbits that result from the adopted distance from the Sun and corresponding 
uncertainties (Sect.~\ref{intro}).
The interesting fact is that the orbit of Stephenson\,2 passes close to the impact site of
$\omega$\,Centauri at a comparable time, within  uncertainties (see below). Since Stephenson\,2 and
BDSB\,122 have almost the same position (within uncertainties), the same conclusions hold for the
latter cluster. The XZ and YZ-plane projections (bottom panels) show that $\omega$\,Centauri emerged
at $\approx45\degr$ from the plane to its present position.

To probe orbital uncertainties owing to the adopted potential, we also employed the potential 
model of \citet{Flynn96} -- hereafter FSC96 -- and tested consequences of variations of $\pm10\%$ 
in the input parameters of JHB96. The results are shown in Fig.~\ref{fig2} (top-right panel), from
which we conclude that orbit variations due to the adopted potential are much smaller than our error 
ellipsoid (Fig.~\ref{fig3}, right panel). 

Close-ups of the $\omega$\,Centauri impact site and the back-traced positions of Stephenson\,2 
are shown in Fig.~\ref{fig3} (left panel) for a Sun's distance to the Galactic centre of 7.2\,kpc 
and 7.6\,kpc. It is clear that the latter value is not critical for the encounter. The right panel 
shows the error ellipsoid of several impact site simulations computed by varying initial conditions 
according to the errors in the different relevant input quantities. The ellipsoid reflects variations
implied by velocity uncertainties in the PM, radial velocity and present position of $\omega$\,Centauri 
along the line of sight in the (U,V,W) velocities. The impact obtained with a Galactocentric distance
$\dgc=7.6$\,kpc is also shown. The disk-orbit of Stephenson\,2 crosses the $\omega$\,Centauri ellipsoid 
error distribution.  The range in impact site to proto-cluster separations contains distances smaller 
than $\approx1$\,kpc, with an average separation of $\sim500$\,pc (intersection area in Fig.~\ref{fig3}, 
right panel). Larger separations would require prohibitive expansion velocities, despite the fact 
that we are dealing with an encounter in a denser, central part of the disk, while in NGC\,4559, the 
event was external. 

For the GC-induced formation hypothesis to be valid, the time-scales associated with the
onset of star formation (after impact), duration of star formation and the cluster age, should
be compatible with the disk-crossing age. Following \citet{PC08}, the first time scale in not 
well known, ranging from virtually instantaneous, i.e. negligible as compared to the cluster age, 
to 15\,Myr (\citealt{LD94}) and 30\,Myr (\citealt{Wallin96}). The star formation time-scale may
be short, $\approx2\times10^5$\,yr, as suggested by \citet{MC02} for stars more massive than 8\,\ms.
Given that the ages of Stephenson\,2 and BDSB\,122 are within 12-17\,Myr and 7-12\,Myr, respectively,
$\omega$\,Centauri, which crossed the disk $\approx24$\,Myr ago, may have triggered their formation
only if the star-formation onset occurred in less than $\approx15$\,Myr, which is within the accepted
range. In the case of NGC\,4559 these time-scales combined are less than $\sim30$\,Myr 
(\citealt{Soria05}). 

The above clues suggest an interesting event, that the most recent crossing of
$\omega$\,Centauri through the disk occurred very close to the sites where two massive OCs
were formed. 
Both Stephenson\,2 and BDSB\,122 are somewhat younger than the age of the impact, and
the differences of a few Myrs are consistent with the shock propagation and subsequent star formation.
The overall evidence gathered in the present analysis supports $\omega$\,Centauri as the origin of
this localised star formation in the Galaxy, which harbours two of the more massive known OCs.

This work suggests a scenario where the disk passage of GCs can generate OCs, massive ones 
in particular, as indicated by the orbit of $\omega$\,Centauri and its impact site. As a consequence,
not all OC formation is induced by the spiral density wave mechanism in spiral arms. 

\section{Summary and conclusions}
\label{Conclu}

Globular clusters orbiting the bulge and halo of the Galaxy cross the disk on average
once every 1\,Myr, and significant physical effects of such events on the disk are expected 
to occur. For instance, the impact of a GC passing through the disk can trigger 
star formation either by accumulation of gas clouds around the impact site or by the 
production of an expanding mechanical wave. Time delays are
expected in both cases because of collapse and fragmentation of molecular clouds before star 
formation. This phenomenon has been recently observed in the galaxy NGC\,4559 (e.g. SCP05). 
If this mechanism operates frequently in the Galaxy, the most massive 
GC $\omega$\,Centauri can be taken as an ideal projectile to prospect the state of its last 
impact site in the disk.

As $\omega$\,Centauri hit the Galactic disk $\approx24$\,Myr ago, a major star-forming event 
appears to have occurred close ($\la1$\,kpc) to  the locus that generated two of the most
massive young OCs known in the Galaxy, BDSB\,122 and Stephenson\,2. We suggest the 
connection of these events in a way similar to the impact and shock wave observed in NGC\,4559 
(SCP05). We use a model of the Galactic potential to integrate the orbit of $\omega$\,Centauri.
As shown in Fig.~\ref{fig3}, when the uncertainties in space velocity, distances, and
potential are considered, the error-distributions of $\omega$\,Centauri impact site and the birth-site of 
Stephenson\,2 overlap. Such overlap suggests a scenario where the disk passage and formation of the 
pair of OCs may be physically connected. Alternatively, the time coincidence may have occurred within 
a separation $\la1$\,kpc. In such case, the expanding bubble scenario such as that in NGC\,4559
would apply. The latter case is more probable, since two clusters have 
been formed.

\citet{Levy2000} performed 2D hydrodynamic simulations to study the impact of GCs on the
Galactic disk in the presence of available gas. They found that the moving GC causes a shock 
wave in the gas that propagates through the disk in a kpc scale, thus producing star formation.
More recently, \citet{PC08} simulated in detail the effects of GC impacts on the disk that
basically confirm \citet{Wallin96} and \citet{Levy2000}, even in the absence of gas at the 
impact site. They found focussing of disk material with compression on a scale of $\sim10$\,pc,
that may subsequently attract gas leading to star formation. The compression increases with 
the GC mass.

At this point an interesting question arises. For a rate of $\sim1$\,GC impact per Myr, it can
be expected a high probability of occurring one GC impact within 1-2\,kpc of any location
within the inner Galaxy in about 10\,Myr. However, the star-formation efficiency of these
events appears to be low, according to \citet{PC08}. Indeed, of the 54 GCs with accurate proper 
motions studied by them, only three appear to be associated to young OCs. Possibly, conditions
like GC mass and impact site properties, such as the availability of molecular gas, temperature 
and density, constrain the star-formation efficiency.

Evidence drawn from the present work suggests that GCs can be progenitors of massive OCs. In 
particular we focused on
$\omega$\,Centauri. Density-wave shocks are possibly not the sole responsible for the 
formation of the more massive OCs, a possibility to be further explored, both theoretically 
and observationally. 

\section*{Acknowledgements}
We thank the anonymous referee for suggestions.
We acknowledge partial support from CNPq (Brazil). 


\end{document}